\documentclass[11pt]{article}
\usepackage[utf8]{inputenc}
\usepackage[T1]{fontenc}
\usepackage{graphicx}
\usepackage{longtable}
\usepackage{wrapfig}
\usepackage{rotating}
\usepackage[normalem]{ulem}
\usepackage{amsmath}
\usepackage{amssymb}
\usepackage{capt-of}
\usepackage{hyperref}
\usepackage[citestyle=authoryear-icomp,bibstyle=authoryear, hyperref=true,backref=true,maxcitenames=3,url=true,backend=bibtex]{biblatex}
\usepackage{multirow}
\usepackage{amsmath}
\usepackage{morefloats}
\usepackage{setspace}
\usepackage{graphicx}
\usepackage{float}
\usepackage{changepage}
\usepackage{setspace}
\author{Matt Brigida\thanks{SUNY Polytechnic Institute, 100 Seymour Rd, Utica NY 13502. Email: matthew.brigida@sunypoly.edu}, Kathleen Maceyka\thanks{SUNY Polytechnic Institute, 100 Seymour Rd, Utica NY 13502.}}
\date{\today}
\title{Hedging Deposit Run Risk Prior to the 2023 Regional Banking Crisis}
\begin{document}

\maketitle
\begin{abstract}
In this analysis we determine factors driving the cross-sectional variation in uninsured deposits during the interest rate raising cycle of 2022--2023.  The goal of our analysis is to determine whether banks proactively managed deposit run risk prior to the hiking cycle which produced the 2023 Regional Banking Crisis. We find evidence that interest rate forward, futures, and swap use affected the change in a bank uninsured deposits over the period. Interest rate option use, however, has no effect on the change in uninsured deposits.  Similarly, bank equity levels were uncorrelated with uninsured deposit changes.  We conclude we find no evidence of banks managing run risk via their balance sheet prior to the 2023 Regional Banking Crisis.
\end{abstract}
\vspace*{1cm}



\pagebreak

While bank failures in the 2008 financial crisis were driven by credit risk, the regional bank failures in 2022--2023 were driven by a confluence of interest rate and liquidity risk.  As interest rates rose, asset values at banks fell, which increases the likelihood that bank assets are insufficient to cover liabilities.  If a given bank fails, insured depositors are made whole by the FDIC, however, uninsured depositors may incur losses.  Thus, while interest rates rose, uninsured depositors generally removed their deposits (the deposits \emph{ran}).  This was particularly true for uninsured depositors of non-systemic regional banks.

The question is did banks with a larger proportion of uninsured deposits anticipate this run risk?  In this analysis we test whether banks anticipated their run risk, and prepared their balance sheet accordingly.  

Modeling the effect of interest rate increases on liquidity risk, \cite{drechsler2023banking} conclude that the only ways to hedge deposit run risk are (1) using contracts with convex payoffs (such as options) and (2) raising equity capital as interest rates rise.  Thus we test for relationships between changes in uninsured deposits, and option use and equity levels.

\textbf{\textbf{Hypotheses:}}

\begin{enumerate}
\item Greater levels of swaptions owned in Q4 2021 implies a larger decline in uninsured deposits.
\item Higher levels of bank equity in Q4 2021, and a larger increase in bank equity from 2021 to 2023, implies a larger decline in uninsured deposits.
\end{enumerate}

We also test for the effect of bank interest rate futures, forward, and swap use on the change in uninsured deposits.  

Hypothesis 2 relates both the overall level of equity in Q4 2021 and the change in equity from 2021 to 2023.  If banks behaved as outlined in \cite{drechsler2023banking} they may \emph{proactively} raise equity levels ahead of a likely interest rate hiking cycle.

Beyond hedging run risk, how banks manage deposits is of particular importance given the large effect deposit management has on bank value.  In fact, \cite{egan2022cross} find that two-thirds of the median bank's value is due to deposit productivity.  Further, they find deposit productivity explains much of the cross-sectional variation in the value of banks.

Bank technology use is another driver of heterogeneity in recent bank deposit outflows.  \cite{benmelech2023bank} found evidence that banks which rely on digital banking, and therefore have low bank density, received large deposit inflows between 2016 and 2022. These same banks, however, had large deposit outflows and steep declines in their stock prices during the banking crisis of 2023.  Variation among banks is likely also driven by bank size.  \cite{d2023deposit} find evidence of differing deposit-pricing depending on bank size.

Our research builds on earlier research which showed when deposit rate increases do not match interest rate increases, deposits begin to run (\cite{drechsler2017deposits}). The \cite{diamond1983bank} model focused on the effect of asset losses on bank runs.  They showed a forced sale of bank loans may cause bank liabilities to exceed assets and lead to a bank run.  Deposits may leave the bank due to competing products as well.  Depositors may switch to money market funds if bank deposit rate increases to match money market rate increases (\cite{xiao2020monetary}).  In particular, \cite{hanson2015banks} find runs of uninsured deposits may have a greater effect on a bank's ability to lend long-term.

Our analysis is organized as follows.  Section 1 below describes our data and methods, and section 2 outlines our results. Section 3 concludes.
\section{Data and Methods}
\label{sec:orgc3fcc92}

Below is a chart of the federal funds rate from 2020 through 2024, which explains our choice of Q4 2021 and Q3 2023. Our explanatory variables, and our starting point for uninsured deposits, are observed when the Federal Funds rate was near 0\%, and had been for multiple quarters. We then take the second observation for uninsured deposits after the increase in the Federal Funds rate in Q3 2023.

\begin{figure}[htbp]
\centering
\includegraphics[width=.9\linewidth]{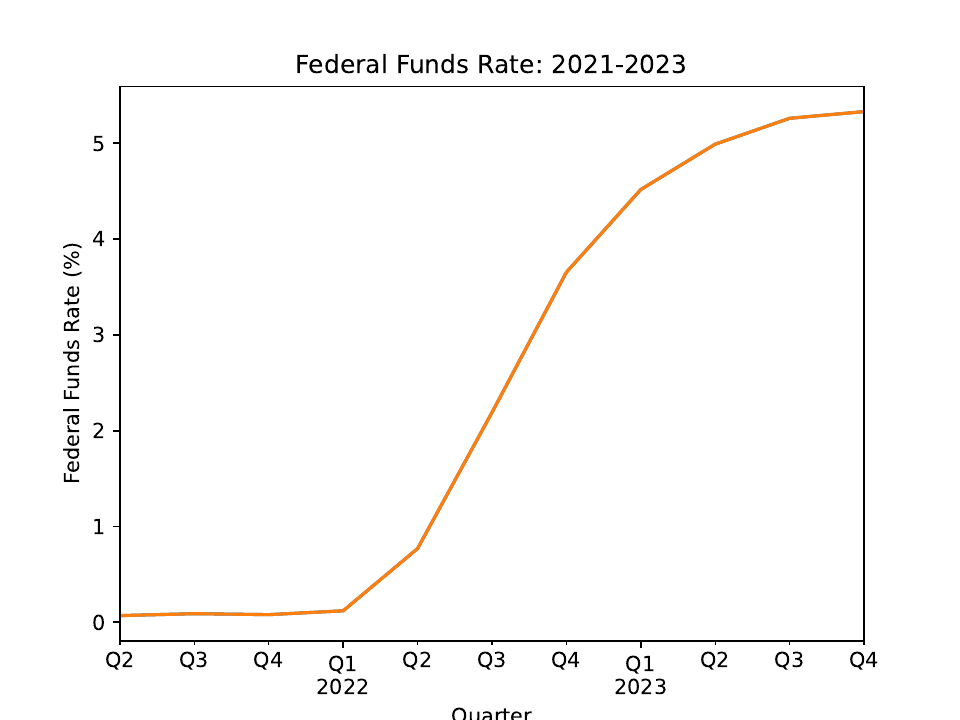}
\caption{The Effective Federal Funds from 6/1/2021 to 12/31/2023.  Data are from the St. Louis Federal Reserve district bank.}
\end{figure}

All data (outside of the Federal Funds rate) are from FDIC Consolidated Reports of Condition and Income (Call Reports).  Banks are required to file call reports quarterly, and they contain comprehensive financial information on bank borrowing, lending, derivative use, as well as financial performance measures.  Table 1 below reports how each variable in our analysis was calculated.

We calculate the change in the percent of uninsured deposits (our dependent variable) using call reports from Q4 2021 and Q3 2023.  For each quarter we calculate percent uninsured deposits as the difference of total deposits (FDIC data code \emph{DEP}) and insured deposits (FDIC data code \emph{DEPINS}) divided by total deposits.  The change in the percent of uninsured deposits is the percent uninsured deposits in Q3 2023 less those of Q4 2021.

Note, we don't show the calculation for average insured, and uninsured, liability repricing due to the length of the calculation. However these variables are time-until-maturity (or repricing) weighted averages of insured and uninsured liabilities.

First note that, as we expect, the mean change in uninsured deposits from Q4 2021 to Q3 2023 is negative.  Specifically, on average uninsured deposits declined by 1.9 percentage points.  Prior to the increase in the federal funds rate the average bank had 27\% of uninsured deposits, whereas after the increase it had 25\%.

Notably, while we have deposit data for 4551 banks, we have interest rate derivative data for roughly 1200 banks.  This is consistent with \cite{jiang2023limited} which showed few banks hedged with interest rate swaps or swaptions.  Not hedging appears to hinder a bank's ability to lend and thus affects the real economy.  \cite{Krainer_2023} find evidence that unhedged firms reduced lending more than firms which hedged due to losses on assets during the 2022 interest rate hiking cycle.

\begin{table}[htbp]
\caption{Variable calculations.  The codes are from the FDIC's Bankfind Application Programming Interface (\url{https://banks.data.fdic.gov/docs/summary\_properties.yaml})}
\centering
\begin{tabular}{ll}
\hline
Variable & Calculation\\
\hline
Uninsured Deposits & (DEP - DEPINS) / DEP\\
Brokered Deposits & BRO / DEP\\
Wholesale funding ratio & (FREPP + NTRTMLG) / ASSET\\
Equity ratio & EQ / ASSET\\
Total assets & ASSET\\
Interest Rate Swaps & RTNVS / ASSET\\
Interest Rate Options Bought & RTPOC / ASSET\\
Interest Rate Options Sold & RTWOC / ASSET\\
\hline
\end{tabular}
\end{table}

\begin{table}[!htbp] \centering 
  \caption{Summary Statistics.  All variables are from Q4 2021 (except Q3 2023 Uninsured Deposits).  Equity is the ratio of Total Equity to Assets.  Initial \% uninsured deposits is uninsured deposits in Q4 2021. log(assets) is the natural logarithm of total assets.  Brokered Dep. is the ratio of brokered to total deposits.  Unin. and Ins. Liab. Repricing is the weighted-average time until uninsured and insured liabilities are repriced respectively.  IR Futures Forwards is the net notional amount of Interest Rate Futures and Forward contracts. IR Swaps is the net notional amount of interest rate swaps.  IR Options Bought and Sold are the net notional amounts of options on interest rate products bought and sold respectively.} 
  \label{} 
\begin{tabular}{@{\extracolsep{5pt}}lccccc} 
\\[-1.8ex]\hline 
\hline \\[-1.8ex] 
Statistic & \multicolumn{1}{c}{N} & \multicolumn{1}{c}{Mean} & \multicolumn{1}{c}{St. Dev.} & \multicolumn{1}{c}{Min} & \multicolumn{1}{c}{Max} \\ 
\hline \\[-1.8ex] 
\% Uninsured Dep. 2021 & 4,553 & 0.270 & 0.149 & 0.00004 & 1.000 \\ 
\% Uninsured Dep. 2023 & 4,551 & 0.250 & 0.135 & 0.002 & 1.000 \\ 
Change Uninsured Dep. & 4,551 & $-$0.019 & 0.076 & $-$0.544 & 0.680 \\ 
Tot IR Contracts & 4,553 & 0.054 & 1.434 & 0.000 & 93.462 \\ 
IR Futures Forwards & 1,172 & 0.025 & 0.314 & 0.000 & 8.503 \\ 
IR Swaps & 1,172 & 0.121 & 1.793 & 0.000 & 58.289 \\ 
IR Options Bought & 1,165 & 0.018 & 0.420 & 0.000 & 14.222 \\ 
IR Options Sold & 1,172 & 0.023 & 0.442 & 0.000 & 14.964 \\ 
Unin. Liab. Repricing & 4,389 & 0.973 & 0.537 & 0.167 & 4.000 \\ 
Insured Liab. Repricing & 4,487 & 1.074 & 0.390 & 0.167 & 3.274 \\ 
Brokered Deposits & 4,544 & 0.015 & 0.043 & 0.000 & 0.605 \\ 
Equity & 4,544 & 0.115 & 0.067 & 0.021 & 0.992 \\
Equity to Uninsured & 4,544 & 10.460 & 517.319 & 0.024 & 34,549.800 \\ 
log(assets) & 4,553 & 12.843 & 1.491 & 8.127 & 21.919 \\ 
\hline \\[-1.8ex] 
\end{tabular} 
\end{table}
\section{Results}
\label{sec:orgd2a015d}

Regression results are in Table 3 below.  We do not find evidence in favor of either of our hypotheses.  The coefficient on equity is positive and significant when the regression does not include variables on derivative use, however is negative and significant when these variables are included.  Note, the inclusion of derivative measures (specifically interest rate option) decreases the sample size from roughly 4500 to 1123 banks.  So it is possible that our hypothesis is supported over all banks, though not for the restricted sample of (mostly large) banks which hedge using derivatives.

The coefficients on interest rate option use are all insignificant. This is evidence against banks employing options to hedge deposit run risk as outlined in \cite{drechsler2023banking}.  That said, FDIC call report data only provides the notional amount of options bought and sold, and not the type of option.  Therefore it is certainly possible a relationship exists, however it is hidden by the lack of specificity of FDIC data.

Across all regressions in Table 3, declines in uninsured deposits were greater if the bank had more uninsured deposits prior to the interest rate increase.  Similarly, uninsured deposit declines were greater if a bank had more brokered deposits.  This makes sense given brokered deposits (being wholesale funding) have a beta near 1 with respect to changes in the short-term interest rate. 

The models without derivative variables explain approximately 20\% of the cross-sectional variation in the change in uninsured deposits. However adding in variables on interest rate derivatives increases the adjusted \(R^2\) to about 31\%, though it reduces our sample size by over 3000 banks.

In the subsample of firms with derivative (futures, forwards, swaps, or options) positions, the greater the notional amount of futures/forwards, the more uninsured deposits declined. Conversely, the more notional in swap contracts, the less the decline in uninsured deposits.  Unfortunately, while we have notional amount for derivative positions, we do not have the net position.  So we, for example, can't tell if a bank is net long or short interest rate futures.

\begin{table}[!htbp] \centering 
  \caption{Determinants of the Change in the Percent of Bank Uninsured Deposits.  Standard errors are heterskedasticity-consistent, and are below the coefficients in parentheses. ****, ***, **, and * denote significance at the 0.1\%, 1\%, 5\%, and 10\% levels respectively.  Our dependent variable is the \% uninsured deposits in Q3 2023 less \% uninsured deposits in Q4 2021.  All explanatory variables are from Q4 2021. Equity is the ratio of Total Equity to Assets.  Initial \% uninsured deposits is uninsured deposits in Q4 2021. log(assets) is the natural logarithm of total assets.  Brokered Dep. is the ratio of brokered to total deposits.  Unin. and Ins. Liab. Repricing is the weighted-average time until uninsured and insured liabilities are repriced respectively.  IR Futures Forwards is the net notional amount of Interest Rate Futures and Forward contracts. IR Swaps is the net notional amount of interest rate swaps.  IR Options Bought and Sold are the net notional amounts of options on interest rate products bought and sold respectively.} 
  \label{} 
\begin{tabular}{@{\extracolsep{5pt}}lcccc} 
\\[-1.8ex]\hline 
\hline \\[-1.8ex] 
 & \multicolumn{4}{c}{\textit{Dependent variable:}} \\ 
\cline{2-5} 
\\[-1.8ex] & \multicolumn{4}{c}{Change in Percent Uninsured Deposits} \\ 
\\[-1.8ex] & (1) & (2) & (3) & (4)\\ 
\hline \\[-1.8ex] 
 Initial \% Uninsured Dep. & $-$0.218$^{****}$ & $-$0.234$^{****}$ & $-$0.308$^{****}$ & $-$0.312$^{****}$ \\ 
  & (0.012) & (0.013) & (0.022) & (0.022) \\ 
  log(assets) & $-$0.001 & $-$0.0003 & 0.004$^{***}$ & 0.003$^{**}$ \\ 
  & (0.001) & (0.001) & (0.001) & (0.001) \\ 
  Brokered Dep. & $-$0.108$^{****}$ & $-$0.098$^{***}$ & $-$0.241$^{****}$ & $-$0.237$^{****}$ \\ 
  & (0.030) & (0.032) & (0.062) & (0.062) \\ 
  Equity & 0.075$^{***}$ & 0.022 & $-$0.097$^{*}$ & $-$0.097$^{*}$ \\ 
  & (0.016) & (0.036) & (0.056) & (0.057) \\ 
  Unin.\ Liab.\ Repricing &  & $-$0.001 & 0.0004 & 0.001 \\ 
  &  & (0.002) & (0.006) & (0.005) \\ 
  Insured\ Liab.\ Repricing &  & 0.0001 & 0.005 & 0.005 \\ 
  &  & (0.003) & (0.008) & (0.007) \\ 
  IR\ Futures\ Forwards &  &  & $-$0.026$^{***}$ &  \\ 
  &  &  & (0.002) &  \\ 
  IR Swaps &  &  &  & 0.009$^{**}$ \\ 
  &  &  &  & (0.004) \\ 
  IR\ Options\ Bought &  &  & 0.009 & $-$0.006 \\ 
  &  &  & (0.049) & (0.052) \\ 
  IR\ Options\ Sold &  &  & 0.005 & $-$0.028 \\ 
  &  &  & (0.046) & (0.048) \\ 
  Constant & 0.049$^{***}$ & 0.048$^{***}$ & 0.022 & 0.032 \\ 
  & (0.010) & (0.011) & (0.020) & (0.019) \\ 
 \hline \\[-1.8ex] 
Observations & 4,544 & 4,372 & 1,123 & 1,123 \\ 
R$^{2}$ & 0.196 & 0.198 & 0.313 & 0.309 \\ 
Adjusted R$^{2}$ & 0.195 & 0.197 & 0.307 & 0.303 \\ 
\hline 
\hline \\[-1.8ex] 
\textit{Note:}  & \multicolumn{4}{r}{$^{*}$p$<$0.1; $^{**}$p$<$0.05; $^{***}$p$<$0.01} \\ 
\end{tabular} 
\end{table}

\vspace*{1cm}

\noindent \emph{Including the change in equity}\\

\cite{drechsler2023banking} determine one of the few ways banks can hedge run risk is to increase equity as the short rate increases.  We therefore include the change in total equity over the interest rate hiking cycle in our regressions (Table 4).  The coefficient of total equity is insignificant however.

\begin{table}[!htbp] \centering 
  \caption{Determinants of the Change in the Percent of Bank Uninsured Deposits.  Standard errors are heterskedasticity-consistent, and are below the coefficients in parentheses. ****, ***, **, and * denote significance at the 0.1\%, 1\%, 5\%, and 10\% levels respectively. Our dependent variable is the \% uninsured deposits in Q3 2023 less \% uninsured deposits in Q4 2021.  All explanatory variables are from Q4 2021 except for the Change in Equity. Equity is the ratio of Total Equity to Assets.  Change in Equity is Total Equity to Assets in Q3 2023 less Total Equity to Assets in Q4 2021.  Initial \% uninsured deposits is uninsured deposits in Q4 2021. log(assets) is the natural logarithm of total assets.  Brokered Dep. is the ratio of brokered to total deposits.  Unin. and Ins. Liab. Repricing is the weighted-average time until uninsured and insured liabilities are repriced respectively.  IR Futures Forwards is the net notional amount of Interest Rate Futures and Forward contracts. IR Swaps is the net notional amount of interest rate swaps.  IR Options Bought and Sold are the net notional amounts of options on interest rate products bought and sold respectively.} 
  \label{} 
\begin{tabular}{@{\extracolsep{5pt}}lcccc} 
\\[-1.8ex]\hline 
\hline \\[-1.8ex] 
 & \multicolumn{4}{c}{\textit{Dependent variable:}} \\ 
\cline{2-5} 
\\[-1.8ex] & \multicolumn{4}{c}{Change in Percent Uninsured Deposits} \\ 
\\[-1.8ex] & (1) & (2) & (3) & (4)\\ 
\hline \\[-1.8ex] 
 Initial \% Uninsured Dep. & $-$0.225$^{***}$ & $-$0.237$^{***}$ & $-$0.317$^{***}$ & $-$0.321$^{***}$ \\ 
  & (0.008) & (0.008) & (0.016) & (0.016) \\ 
  log(assets) & $-$0.001 & $-$0.00002 & 0.004$^{***}$ & 0.004$^{***}$ \\ 
  & (0.001) & (0.001) & (0.001) & (0.001) \\ 
  Brokered Dep. & $-$0.101$^{***}$ & $-$0.099$^{***}$ & $-$0.220$^{***}$ & $-$0.216$^{***}$ \\ 
  & (0.022) & (0.024) & (0.042) & (0.042) \\ 
  Equity & 0.071$^{***}$ & 0.016 & $-$0.104$^{*}$ & $-$0.102 \\ 
  & (0.016) & (0.027) & (0.062) & (0.062) \\ 
  Change\ in\ Equity & $-$0.015 & 0.031 & 0.061 & 0.056 \\ 
  & (0.028) & (0.033) & (0.069) & (0.069) \\ 
  Unin.\ Liab.\ Repricing &  & $-$0.001 & 0.001 & 0.001 \\ 
  &  & (0.002) & (0.005) & (0.005) \\ 
  Ins.\ Liab.\ Repricing &  & 0.0002 & 0.005 & 0.005 \\ 
  &  & (0.003) & (0.007) & (0.007) \\ 
  IR\ Futures\ Forwards &  &  & $-$0.027$^{***}$ &  \\ 
  &  &  & (0.009) &  \\ 
  IR\ Swaps &  &  &  & 0.010$^{*}$ \\ 
  &  &  &  & (0.005) \\ 
  IR\ Options\ Bought &  &  & 0.008 & $-$0.006 \\ 
  &  &  & (0.039) & (0.042) \\ 
  IR\ Options\ Sold &  &  & 0.006 & $-$0.029 \\ 
  &  &  & (0.038) & (0.037) \\ 
  Constant & 0.048$^{***}$ & 0.046$^{***}$ & 0.019 & 0.029 \\ 
  & (0.010) & (0.011) & (0.021) & (0.022) \\ 
 \hline \\[-1.8ex] 
Observations & 4,566 & 4,381 & 1,129 & 1,129 \\ 
R$^{2}$ & 0.199 & 0.200 & 0.319 & 0.316 \\ 
Adjusted R$^{2}$ & 0.199 & 0.199 & 0.313 & 0.309 \\ 
\hline 
\hline \\[-1.8ex] 
\textit{Note:}  & \multicolumn{4}{r}{$^{*}$p$<$0.1; $^{**}$p$<$0.05; $^{***}$p$<$0.01} \\ 
\end{tabular} 
\end{table}
\section{Conclusion}
\label{sec:org2343937}

Previous theoretical research (\cite{drechsler2023banking}) determined that the only ways to hedge uninsured deposit run risk during a rate hiking cycle was to use derivatives with convex payoffs such as options, and to increase equity along with the short rate.  Using bank-level FDIC Call report data we tested whether there was a relationship between the change in uninsured deposits, and option and equity data, over the 2022--2023 short rate increases. We found no evidence that equity levels, or the change in equity over the hiking cycle, was related to the change in uninsured deposits.  Similarly, we found no evidence that the notional amount interest rate swap options bought or sold have any relationship with the subsequent change in uninsured deposits over the hiking cycle.

The most prominent determinants of a bank's change in uninsured deposits over the hiking cycle was the bank's overall level of uninsured and brokered deposits.  Higher levels of both led to greater declines in uninsured deposits.  Additionally the results may point to interest rate swaps and futures/forwards having an effect on the behavior of uninsured deposits.  However these securities are not an effective hedge against run risk.

In sum, we find no evidence of banks proactively managing run risk.  Uninsured deposit changes during the 2022--2023 rate hiking cycle were purely a function of prior levels of uninsured deposits, and the levels of other brokered deposits which have a beta near 1.  Though, our results are affected by the lack of granularity in FDIC Call Report data on bank option use.

\clearpage

\printbibliography
\end{document}